\documentclass[superscriptaddress,amsmath,amssymb]{revtex4}
\usepackage[T1]{fontenc}
\usepackage{graphicx}
\usepackage{dcolumn}
\usepackage{bm}
\usepackage{array}

\begin{document}
\title{Neutron Energy Dependence of Delayed Neutron Yields and its Assessments}
\author{Futoshi Minato}
\affiliation{Nuclear Data Center, Japan Atomic Energy Agency, Tokai 319-1195, Japan}
\affiliation{NSCL/FRIB Laboratory, Michigan State University, East Lansing, Michigan 48824, USA}

\date{\today}

\begin{abstract}
Incident neutron energy dependence of delayed neutron yields of uranium and plutonium isotopes is investigated. A summation calculation of decay and fission yield data is employed, and the energy dependence of the latter part is considered in a phenomenological way. Our calculation systematically reproduces the energy dependence of delayed neutron yields by introducing an energy dependence of the most probable charge and the odd-even effect. The calculated fission yields are assessed by comparison with JENDL/FPY-2011, delayed neutron activities, and decay heats. Although the fission yields in this work are optimized to delayed neutron yields, the calculated decay heats are in good agreement with the experimental data. Comparison of the fission yields calculated in this work and JENDL/FPY-2011 gave an important insight for the evaluation of the next JENDL nuclear data. 
\end{abstract}

\maketitle

\section{Introduction}
\label{intro}
Some fission fragments located on neutron-rich side of the nuclear chart emit neutrons after the $\beta$-decay, which are the so-called delayed neutrons. Delayed neutron yields resulting from neutron induced fission of actinide nuclides are crucial to making a light water reactor controllable because they lengthen the reactor period to a time scale long enough to keep the critical state of the reactor. Delayed neutrons play a meaningful role in other research fields as well. One of the examples is $r$-process nucleosynthesis, which is the leading candidate for producing heavy elements by successive neutron capture at an astronomical site. At the ``freeze-out" phase where environmental neutrons are exhausted in the astronomical sites, nuclides produced by the $r$-process start going towards the $\beta$-stability line by $\beta$-decay. Then, the delayed neutron emission from some nuclei accompanied by $\beta$-decay shifts the abundance pattern of the elements by re-accumulation of the neutron density \cite{Arcones2011}. Delayed neutrons also have potential to be applied to the interrogation of actinides for homeland security \cite{Lakosi2008, Sari2013}.

The delayed neutron yield can be calculated by a summation calculation of $\beta$-decay and fission yield data. The latter depends on incident neutron energy, and delayed neutron yields therefore change along with excitation energies. It usually increases moderately up to incident neutron energy around $E_n=3$ MeV and decreases rapidly from about $4$ MeV as the second chance fission occurs \cite{Ohsawa2001}. These phenomena were measured actively by the Obninsk group in the past twenty years \cite{Piksaikin1997, Roshchenko2006b, Roshchenko2006b2, Piksaikin2006, Piksaikin2013}. From a theoretical point of view, Alexander and Krick studied the energy dependence by using a summation calculation about forty years ago \cite{Alexander1977}, in which fission yields were calculated by Wahl's approach \cite{Wahl} and 47 precursors were considered. Taking into account the odd-even effect on the most probable charge, its energy dependence, and multi-chance fission, they were able to show a reasonable agreement with the experimental data of $^{235}$U. However, they did not investigate other actinides and higher energy than $E_n=15$ MeV. In a recent work, a similar calculation for $^{235}$U was performed up to $E_n=8$ MeV by Ohsawa and Miura \cite{Ohsawa2001}, and Ohsawa and Fukuda \cite{Ohsawa2007}. They considered a highly excited state of primary fission fragments, which facilitates multiple neutron emission, in order to explain the energy dependence of delayed neutron yields between $4$ to $7$ MeV. Ohsawa and Hambsch have also discussed delayed neutron yields around the resonance region for $^{235}$U and $^{239}$Pu by using the multimodal random neck rupture model \cite{Ohsawa2004}.

Aiming at a high secure design of next generation nuclear reactors and a reduction of radioactive wastes including minor actinides, a variety of nuclear reactors using different fuel compositions and different neutron energies from light water reactors have been designed. Typical examples are accelerator-driven systems and fast breeder reactors. In order to design them reliably, nuclear data (especially neutron-nucleus reaction cross sections) have been updated. However, the evaluated data of delayed neutron yields have not been renewed as actively as other nuclear data. For this reason, a new international activity concerning delayed neutrons, which aims to produce new evaluated data, is now in progress \cite{IAEACRP}.

The present evaluated delayed neutron data compiled in the Japanese evaluated nuclear data library (JENDL-4.0) \cite{JENDL} are performed by an interpolation of existing data or systematics proposed by Tuttle \cite{Tuttle1979}, Benedetti \cite{Benedetti1982}, Waldo \cite{Waldo1981}, and Manero \cite{Manero1972}. As a consequence, they are usually given in the form of structure-less lines, which are not natural indeed. Ideally, the evaluated delayed neutron data should be consistent with the summation calculation performed with the corresponding evaluated decay and fission yield data. However, the quality of the summation calculation using them is not accurate enough, so that delayed neutron yields have always been evaluated independently.

In this work, we present a method for evaluating the incident neutron energy dependence of delayed neutron yields using least square fitting. To this end, a function (or model) to be used for the fitting has to be carefully chosen. Fortunately, several developments have been achieved in theory as well as experiment since Alexander and Krick had studied the energy dependence of delayed neutron yields. Those developments, including Ohsawa's idea \cite{Ohsawa2007}, are able to improve the quality of the summation calculation and the reliability of fitting.

We discuss the incident neutron energy dependence of the delayed neutron yield of uranium ($A=233, 235, 236, 238$) and plutonium isotopes ($A=239-242$) in a similar way to Alexander and Krick's approach. A new parametrization of the energy dependence of fission yields is introduced and the parameters are determined by least square fitting to the available experimental data of delayed neutron yields. The calculated fission yields are compared with the evaluated fission yield data of JENDL/FPY-2011 \cite{Katakura2011}. It should be noted that fission yield data affects not only delayed neutrons but also decay heat. To assess the present summation calculation, we therefore take care of decay heat as well as delayed neutron activities.

The content of this paper is the following. Section II describes our calculation method and Sec. III discusses the result of the energy dependence of delayed neutron yields. In sec. IV, fission yields obtained in this work are compared with JENDL/FPY-2011 \cite{Katakura2011}. We also show delayed neutron yields and decay heats calculated by the summation calculation and compare them with experimental data. In sec. V, conclusions of the present work and future plans are given.

\section{Calculation}
\label{calculation}
\subsection{Fission Product Yields}
Delayed neutron yields $\nu(E_n)$ as a function of incident neutron energy $E_n$ in this work are calculated by the summation calculation
\begin{equation}
\begin{split}
\nu(E_n)&=\int_0^\infty \nu_t(E_n,t)dt\\
\nu_t(E_n,t)&=\sum_i^{\text{all FP}} P_{ni}\lambda_i y_i(E_n,t)
\end{split}
\label{delayedneutron}
\end{equation}
where $y_i(E_n,t)$ is the fission yields of nuclide $i=(A,Z,M)$ at time $t$ which explicitly have incident neutron energy dependence, and $\lambda_i$ and $P_{ni}$ are the decay constant and the delayed neutron branching ratios of nuclide $i$, respectively. Here, the symbol $A$, $Z$, and $M$ stand for mass number, atomic number and isomeric states of fission fragments, respectively. We assume that fission occurs at $t=0$.

The fission yields at time $t$ are calculated by the equation
\begin{equation}
\frac{dy_i}{dt}(E_n,t)=-\lambda_iy_i(E_n,t)+\sum_{j\ne i}b_{j\rightarrow i}\lambda_jy_j(E_n,t)
\end{equation}
with the initial condition defined as
\begin{equation}
y_i(E_n,t=0)\equiv y(A,Z,M,E_n)=Y(A,E_n)f(A,Z,E_n)r_M(A,Z,E_n),
\end{equation}
where $y$ are the independent fission yields, $b_{j\rightarrow i}$ the branching ratio from nuclide $j$ to $i$, $Y(A,E_n)$ the mass yields, $f(A,Z,E_n)$ the fractional independent yields, and $r_M(A,Z,E_n)$ the isomeric yield ratios. In what follows, we omit $E_n$ for simplicity. 
The fractional independent yield is calculated by the so-called $Z_p$ model~\cite{Wahl,Katakura2011} as
\begin{equation}
f(A,Z)=\frac{N(A)}{\sigma_A\sqrt{2\pi}}\int_{Z-0.5}^{Z+0.5}\exp\left(-\frac{(z-Z_p(A)^2}{2\sigma_A^2}\right)dz,
\label{fracindep}
\end{equation}
where $Z_p(A)\equiv Z_p(A,E_n)$ is the most probable charge, $\sigma_A$ the width representing charge distribution, and $N(A)\equiv N(A,E_n)$ the normalization factor determined by
\begin{equation}
\sum_{Z=0}^{\infty}f(A,Z)=1.0.
\end{equation}
The function $f(A,Z)$ gives a global shape of nuclear charge distribution. However, there exist deviations from it, which will be discussed in the sec. \ref{sec.oddeveneffect}

In the work of Alexander and Krick~\cite{Alexander1977}, Wahl's compilations~\cite{Wahl} were used for the mass yield and $Z_p(A)$, and $\sigma_A$ is set to be $0.56 \pm 0.06$. For the energy dependence of $Z_p(A)$, Nethaway's systematics~\cite{Nethaway1974} was used. They considered 47 nuclides as the precursors contributing the delayed neutron yield of $^{235}$U and didn't take into account isomers, even though some of them give a large contribution, like $^{98m, 100m}$Y~\cite{Minato2015}.

In this work, the mass yield $Y(A)$ are calculated by the means of five components of the Gaussian function, in which Katakura's systematics and parameters are used~\cite{Katakura2003}. This systematics takes into account the energy dependence of prompt neutrons, and substantially includes Ohsawa's idea that a highly excited state of primary fission fragments facilitates prompt neutron emissions~\cite{Ohsawa2001, Ohsawa2007}. 
We use the evaluated values of England and Rider for $\sigma_A$~\cite{England} and assume that the most probable charge $Z_p$ can be expressed by the following equations;
\begin{equation}
Z_p(A,E_n)=\alpha Z_p^{0}(A)+\beta c(A)E_n+\gamma E_n^2.
\label{energdep}
\end{equation}
This function is based on the Nethaway's systematics~\cite{Nethaway1974}. The first term of Eq.~\eqref{energdep} is independent of incident neutron energy. The second and third terms in Eq.~\eqref{energdep} represent incident neutron energy dependence. The function $c(A)$ accounts for the fact that the energy dependence of $Z_p$ smoothly changes for light fragments and relatively rapidly for heavy fragments~\cite{Nethaway1974, Roshchenko2006}. The third term proportional to $E_n^2$ is a higher order correction which was not taken into account in~\cite{Nethaway1974}. The parameter $Z_p^0(A)$ is taken from the evaluated values of England and Rider~\cite{England}. 
The values of $\alpha, \beta$ and $\gamma$ (and $\xi$ in sec.~\ref{sec.oddeveneffect}) are determined by the least square fitting to experimental data for $^{233,235,236,238}$U and $^{239-242}$Pu.

The isomeric yield ratios $r_M(A,Z)$ are calculated in the same way as Ref.~\cite{Madland1976, Katakura2011}, in which the angular momentum of the primary fragment $J_{rms}=\sqrt{\langle J^2 \rangle}$ is given at thermal neutron energy, fast neutron energy and $14$ MeV. We approximate the energy dependence of it by linearly interpolating the values at thermal energy and 14 MeV and extrapolating above 14 MeV.

The delayed neutron branching ratios $P_{ni}$ are referred from JENDL/FPD-2011~\cite{Katakura2011} with some corrections reflecting new experimental data~\cite{Miernik2013, Miernik2013b, Korgul2013, Gomez-Hornillos2014, Birch2015, Korgul2015, Alshudifat2016, Miernik2016, Madurga2016}, which were obtained after the release of JENDL/FPD-2011.

\subsection{Odd-Even Effect}
\label{sec.oddeveneffect}
Equation~\eqref{fracindep} gives a global shape of the nuclear charge distribution, however, one can observe differences from it in experimental data~\cite{Roshchenko2006, Amiel1975, Amiel1977}. In particular, the contributions due to odd-mass nuclides systematically tends to be smaller than the value given by Eq.~\eqref{fracindep} while that due to even-mass nuclides tends to be larger. This odd-even effect can be attributed to the difference of binding energy of odd and even nuclides, which is substantially triggered by the pairing correlation.

It was shown by Alexander and Krick~\cite{Alexander1977} that the odd-even effect in the charge distribution has a significant influence on the delayed neutron yields. To account for it, they simply multiplied a factor $(1+C)$ for even-$Z$ mass and $(1-C)$ for odd-$Z$ mass by the fission yields~\cite{Alexander1977}, where the factor $C$ depends on the incident neutron energy.

To take into account the odd-even effect in this work, the method proposed by Madland and England ~\cite{Madland1976b} was adopted, where $(1+C_{ee})$, $(1-C_{oo})$, $(1-C_{oe})$, and $(1+C_{eo})$ are multiplied to the fission yields of even-even, odd-odd, odd-even, and even-odd nuclides, respectively. We set $C_{ee}=C_{oo}$ and $C_{oe}=C_{eo}$ in this work. The factors $C_{ee}$ and $C_{oe}$ correspond to the factors $X+Y$ and $X-Y$ defined in Ref.~ \cite{Madland1976b}, respectively. The renormalization is carried out after multiplying the factors.

In general, the pairing correlation decreases as the temperature of nuclide increases (or, in other words, the excitation energy of the compound state), and it vanishes at a certain temperature due to the phase transition from super-fluid to normal-fluid states. The factor of the odd-even effect would have a similar temperature dependence. Alexander and Krick adopted a linear extrapolation of $C=25\%$ at thermal neutron energy and $C=10\%$ at $E_n=2$ MeV, $0\%$ from $E_n=3.33$ to $15$ MeV for $^{235}$U based on experimental results (see Ref.~\cite{Alexander1977} and references therein). In this work, we approximate the energy dependence of $C_{ee}$ and $C_{oe}$ as
\begin{equation}
\begin{split}
C_{ee}(E_n)&=C_{ee}^0\frac{2}{1+\exp\left(\xi E_n\right)}\\
C_{oe}(E_n)&=C_{oe}^0\frac{2}{1+\exp\left(\xi E_n\right)},
\end{split}
\label{pairing}
\end{equation}
where $C_{oo}^0$ and $C_{oe}^0$ are taken from Madland and England's evaluation~\cite{Madland1976b} and are listed in Table~\ref{Cfactors}. 
The functions simulate the energy dependence of pairing studied by a microscopic framework~\cite{Kaneko2004}. The value of $\xi$ is determined by the least square fitting along with $\alpha$, $\beta$ and $\gamma$ in Eq.~ \eqref{energdep}.

\begin{table}
\caption{Odd-even factors $C_{ee}^0$ and $C_{oe}^0$ taken from Madland and England's evaluation~\cite{Madland1976b}.}
\begin{tabular}{c|ccccccccccc}
   & $^{233}$U & $^{235}$U & $^{236}$U & $^{238}$U & $^{239}$Pu & $^{240}$Pu & $^{241}$Pu & $^{242}$Pu \\
\hline
$C_{ee}^0$ & 0.341 & 0.344 & 0.270 & 0.100 & 0.190 & 0.210 & 0.100 & 0.140 \\
$C_{oe}^0$ & 0.197 & 0.202 & 0.170 & 0.060 & 0.098 & 0.130 & 0.100 & 0.080
\end{tabular}
\label{Cfactors}
\end{table}

\subsection{Second and Third Chance Fission}
\label{sec.multichancefission}

As incident neutron energy increases, the neutron emission channel starts to compete with the fission one. In such a situation, contributions of second and third chance fission become important for delayed neutron yields. We take them into account in the following  manner;
\begin{equation}
\nu_d(E_x=E_n+S_n)=\nu_{0n}(E_x)\frac{\sigma(n,f)}{\sigma_f}+\nu_{1n}(E_x')\frac{\sigma(n,n'f)}{\sigma_f}+\nu_{2n}(E_x'')\frac{\sigma(n,2nf)}{\sigma_f},
\label{mcfission}
\end{equation}
where $E_x'=E_x-S_{n}-\epsilon_n'$ and $E_x''=E_x-S_{2n}-\epsilon_n'-\epsilon_n''$ are the excitation energies after one and two neutron emission from the compound nucleus ($A,Z$), respectively, $S_n$ and $S_{2n}$ are the one and two neutron separation energies of compound nuclei, respectively, and $\epsilon_n'$ and $\epsilon_n''$ are the kinetic energy of the emitted neutron from $(A,Z)$ and  $(A-1,Z)$ nuclei, respectively. The values $\nu_{0n}, \nu_{1n},$ and $\nu_{2n}$ are the delayed neutron yields calculated by Eq.~\eqref{delayedneutron} of target nuclides $(A,Z), (A-1,Z)$, and $(A-2,Z)$, respectively. The total fission cross section is $\sigma_f=\sigma(n,f)+\sigma(n,n'f)+\sigma(n,2nf)$; $\sigma(n,f)$, $\sigma(n,n'f),$ and $\sigma(n,2nf)$ are the first, second, and third chance fission cross sections, respectively, and we adopted the values of JENDL-4.0~\cite{JENDL} for them. We checked that the contribution of fourth chance fission is negligible below $20$ MeV.

Because the pre-fission neutron spectra of second and third chance fissions are not provided in JENDL-4.0, we simply approximate $\epsilon_n'=\epsilon_n''=0$ to calculate Eq.~\eqref{mcfission}. The effect of finite $\epsilon_n'$ and $\epsilon_n''$ is discussed in sec.~\ref{dny2}.

$^{231,232,234,237}$U and $^{237,238}$Pu may become fissioning nuclides as a consequence of the second or third chance fission of $^{233,234,235,238}$U and $^{239}$Pu. Those delayed neutron yields (except $^{237}$Pu) are taken from JENDL-4.0~\cite{JENDL}. The value of $^{237}$Pu are taken from the systematics of Tuttle~\cite{Tuttle1979} because the adopted value of JENDL-4.0 is too small as compared to other systematics~\cite{Tuttle1979, Benedetti1982, Waldo1981}.

\section{Delayed Neutron Yields}
\subsection{Incident Neutron Energy Dependence of Delayed Neutron Yields}
\label{dny}
Using the formalisms described in the previous section, the delayed neutron yields of uranium and plutonium isotopes are calculated. The parameters are determined by least square fitting to experimental data. We begin with calculating light nuclides, that is to say $^{233}$U and $^{239}$Pu. We subtract the contribution of the second and third chance fissions beforehand from the experimental data to be used for the fitting. We repeat in the same manner when calculating the heavier nuclides as well. In this way, we take into account the contribution of the second and third chance fissions consistently. We summarize the experimental data used for the least square fitting in Table~\ref{experimentaldata}. We used Tuttle's adjusted values (Table 1 of Ref.~\cite{Tuttle1979}) if they are available, but we omitted the data of McGarry et al. for $^{233,235}$U, Sun et al. for $^{238}$U, Shpakov et al. for $^{239}$Pu because they apparently deviate from empirically expected values.

For $^{240,241,242}$Pu, we set the parameter $\gamma$ to be zero because there are less experimental data than the other isotopes and we could not determine it uniquely. The least square fitting was thus performed only on $\alpha, \beta$, and $\xi$.

The obtained parameters are listed in Table \ref{parameters}. The parameter $\alpha$ is uniquely determined around $1$. On the other hand, the parameters $\beta, \gamma$ and $\xi$ relevant to the energy dependence show a strong dependence on nuclides.

\begin{table}
\caption{Experimental data used for the least square fitting.}
\begin{tabular}{c|l}
Nucleus & References \\
\hline
$^{233}$U & Tuttle~\cite{Tuttle1979}, Borzakov et al.~\cite{Borzakov2000}, Piksaikin et al.~\cite{Piksaikin2006}, Cesana et al.~\cite{Cesana1980}\\
$^{235}$U & Tuttle~\cite{Tuttle1979}, Reeder et al.~\cite{Reeder1983}, Saleh et al.~\cite{Saleh1997}, Roshchenko et al.~\cite{Roshchenko1999}, Piksaikin et al.~\cite{Piksaikin1997}\\
$^{236}$U & Roshchenko et al.~\cite{Roshchenko2006}, Bobkov et al.~\cite{Bobkov1989}\\
$^{238}$U & Tuttle~\cite{Tuttle1979}, Piksaikin et al.~\cite{Piksaikin2007}, Meadows~\cite{Meadows1976}\\
$^{239}$Pu & Tuttle~\cite{Tuttle1979}, Piksaikin et al.~\cite{Piksaikin2006}\\
$^{240}$Pu & Tuttle~\cite{Tuttle1979}, Benedetti et al.~\cite{Benedetti1982}, Cesana et al.~\cite{Cesana1980}\\
$^{241}$Pu & Tuttle~\cite{Tuttle1979}, Meadows~\cite{Meadows1976}, Benedetti et al.~\cite{Benedetti1982}, Cesana et al.~\cite{Cesana1980}\\
$^{242}$Pu & Krick and Evans~\cite{Krick1970}, Bobkov et al~\cite{Bobkov1989}\\
\end{tabular}
\label{experimentaldata}
\end{table}

\begin{figure}
\includegraphics[width=0.40\linewidth]{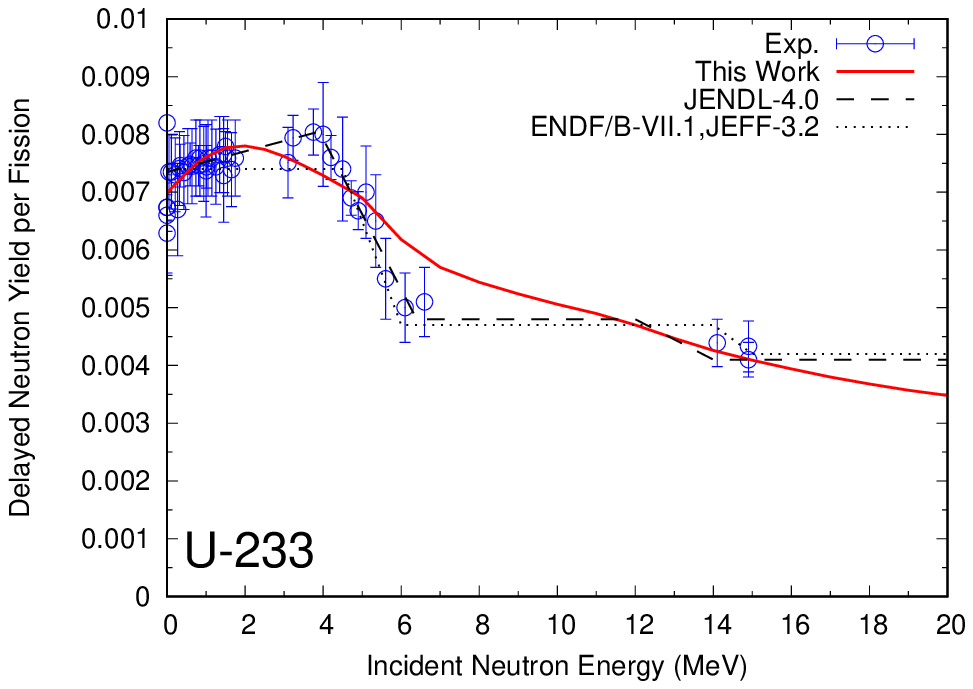}
\includegraphics[width=0.40\linewidth]{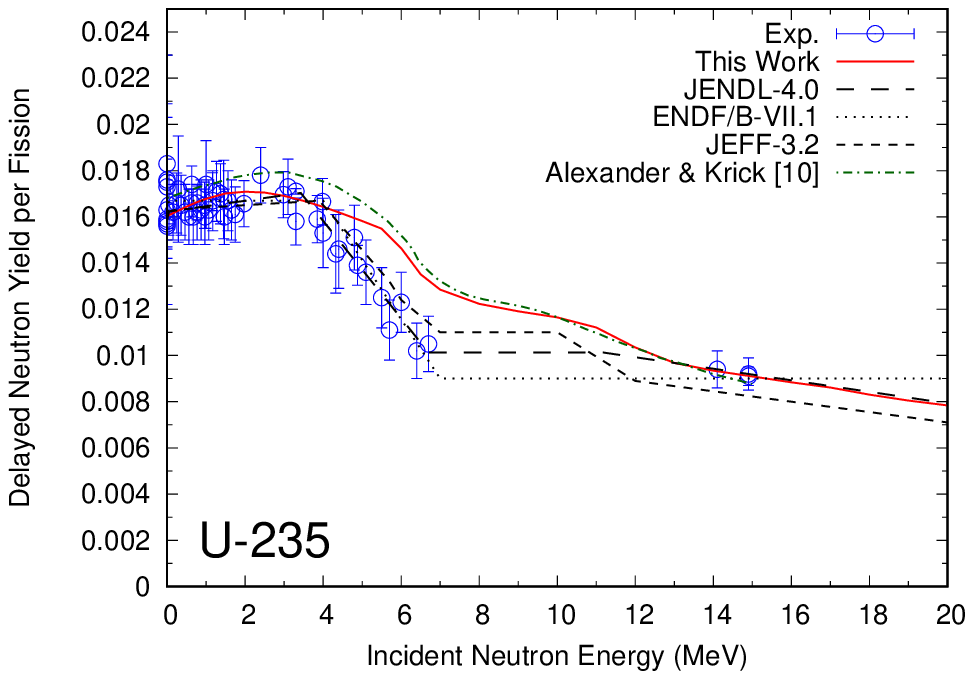}

\includegraphics[width=0.40\linewidth]{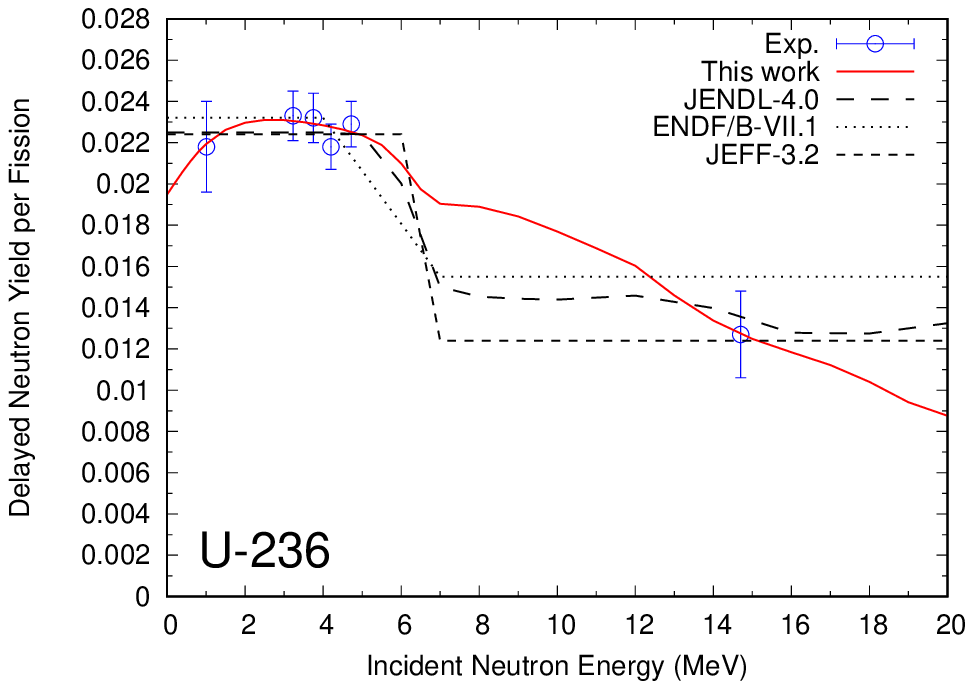}
\includegraphics[width=0.40\linewidth]{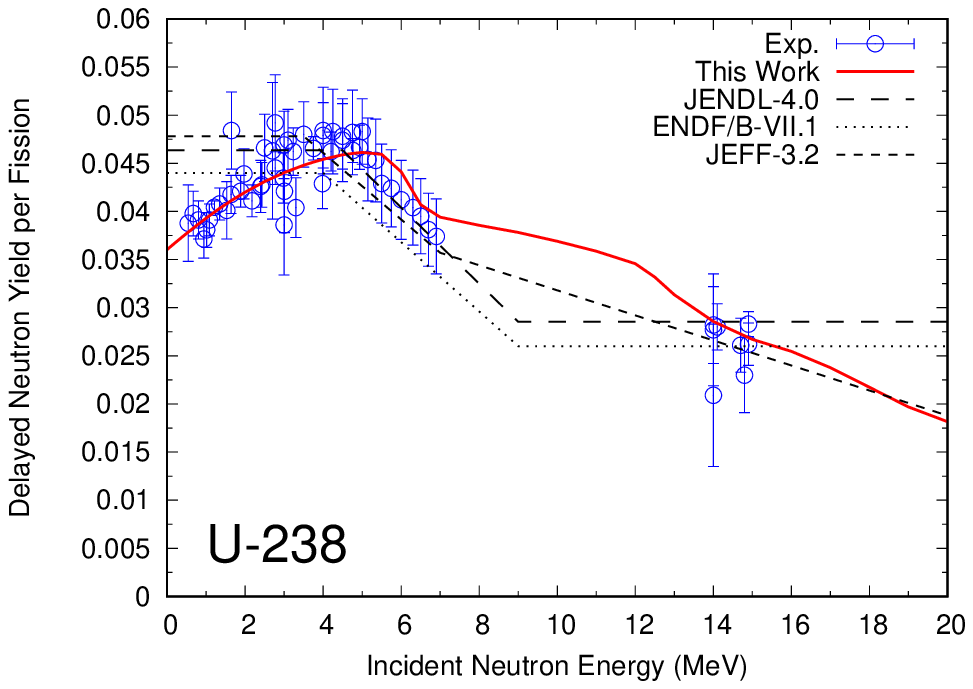}
\caption{Delayed neutron yield of uranium isotopes from $A=233$ to $238$. For $^{235}$U, the median of the result of Alexander and Krick (Fig. 6 of Ref.~\cite{Alexander1977})  is indicated by the dotted-dashed line.}
\label{dnyuran}
\end{figure}

The delayed neutron yields for uranium isotopes of $A=233,235,236,238$ are shown in Fig.~\ref{dnyuran} together with the experimental data and the evaluated values in JENDL-4.0~\cite{JENDL}, ENDF/B-VII.1~\cite{ENDF} and JEFF-3.2~ \cite{JEFF} libraries. For $^{235}$U, the median of the result of Alexander and Krick \cite{Alexander1977} is indicated by the dotted-dashed line. Compared with the experimental data, the energy dependences of $^{233,235,236,238}$U are reproduced reasonably well with our parameter fittings. For $^{235}$U, our estimation of the delayed neutron yields is closer to the experimental data than the median of Alexander and Krick (Fig. 6 of Ref.~\cite{Alexander1977}) at the incident neutron energy from $0$ to $6$ MeV. The steep decreases starting from about $4$ MeV observed in the experimental data of $^{233,235}$U could not be traced well. However, it is not a disappointing result because we do not know reliable delayed neutron yields of $^{232}$U and $^{234}$U as a fissioning nuclei resulting from pre-fission neutron emission of $^{233}$U and  $^{235}$U, respectively. If we modify the delayed neutron yields of $^{232,234}$U, a better result can be obtained. In addition, there exist uncertainties in the evaluated cross sections for second chance fission of $^{233}$U and $^{235}$U. However, predicting the delayed neutron yields of $^{232}$U and $^{234}$U, and their multi-fission cross sections is beyond the scope of the present work. Further effort is required for a precise evaluation of the delayed neutron yields of those nuclides.

In the low energy region before second chance fission occurs, the delayed neutron yields of some nuclides increase moderately up to $2$ MeV, then starts to decrease. Similar behavior is also observed for some plutonium isotopes as we will see later. This is because, as discussed in sec.~\ref{sec.oddeveneffect} the odd-even effect weakens as the excitation energy increases. These structures are consistent with the discussion given in Alexander and Krick's work~ \cite{Alexander1977}. Above about $4$ MeV, our results show a monotonic decrease of delayed neutron yields as a function of incident neutron energy. For $^{233}$U, the evaluated data are close to this work. For $^{235}$U, this work looks similar to the behavior of JEFF-3.2 and matches with JENDL-4.0 above $14$ MeV. For $^{236}$U, this work estimates a different energy dependence from the evaluated data. In particular, the sharp decrease appearing in the evaluated data at about $6$ MeV was not obtained. For $^{238}$U, the experimental data go upward as incident neutron energy increases in the low energy region. Accordingly, this work also shows a similar increase up to $4$ MeV. However, the evaluated data adopts constant values in this energy region. Above $4$ MeV, this work is close to JEFF-3.2, while JENDL-4.0 and ENDF/B-VII.1 give the delayed neutron yields independent of incident neutron energies above about $8$ MeV.

\begin{table}
\caption{Parameters used in Eq.~\eqref{energdep} determined by the least square fitting to the experimental data (read text for more detail).}
\begin{tabular}{c|crrc}
Nuclide  & $\alpha$ & \multicolumn{1}{c}{$\beta$} & \multicolumn{1}{c}{$\gamma$} & $\xi$ \\
\hline
U-$233$& $1.00363$ & $  0.722$ & $1.208\times 10^{-4}$ & $1.159$ \\
U-$235$& $0.99980$ & $ 0.929$ & $-3.561\times 10^{-4}$ & $0.830$ \\
U-$236$& $1.00146$ & $-0.203$ & $1.261\times 10^{-3}$ & $1.402$ \\
U-$238$& $1.00384$ & $-1.800$ & $3.608\times 10^{-3}$ & $0.940$ \\
\hline
Pu-$239$& $0.99997$ & $-0.116$ & $6.816\times 10^{-4}$ & $0.481$ \\
Pu-$240$& $1.00149$ & $0.400$ & 0 & $0.475$ \\
Pu-$241$& $0.99921$ & $0.769$ & 0 & $0.710$ \\
Pu-$242$& $1.00350$ & $0.734$ & 0 & $1.608$ \\
\end{tabular}
\label{parameters}
\end{table}

\begin{figure}
\includegraphics[width=0.40\linewidth]{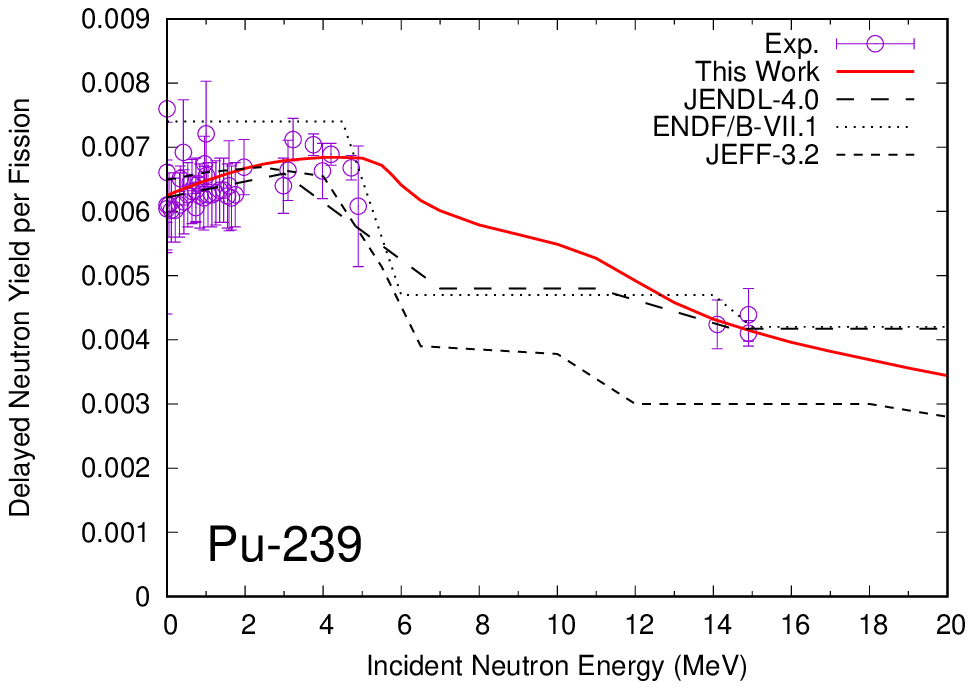}
\includegraphics[width=0.40\linewidth]{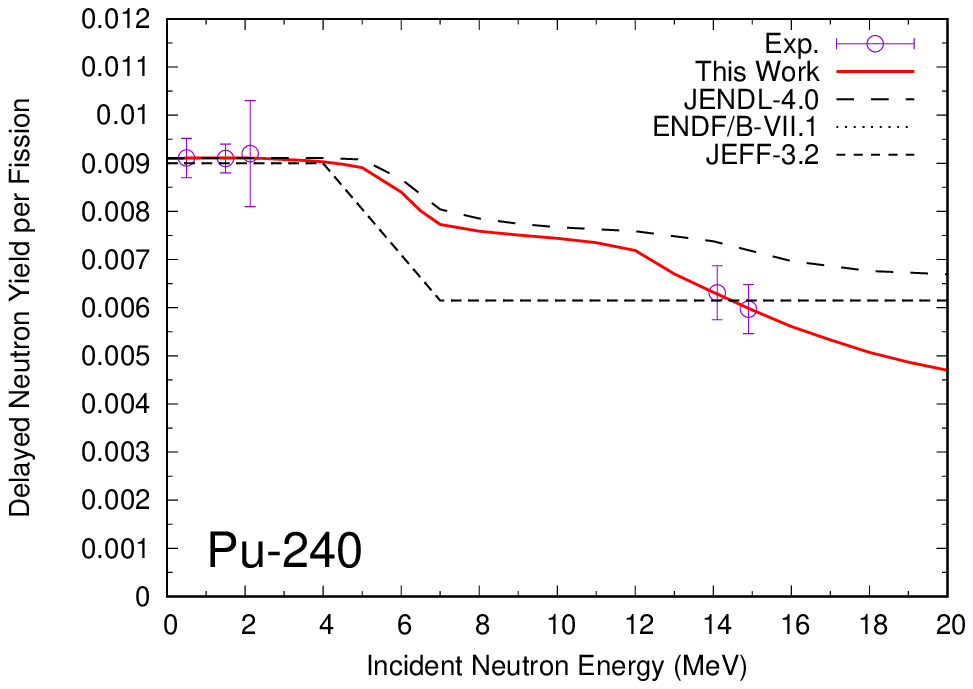}

\includegraphics[width=0.40\linewidth]{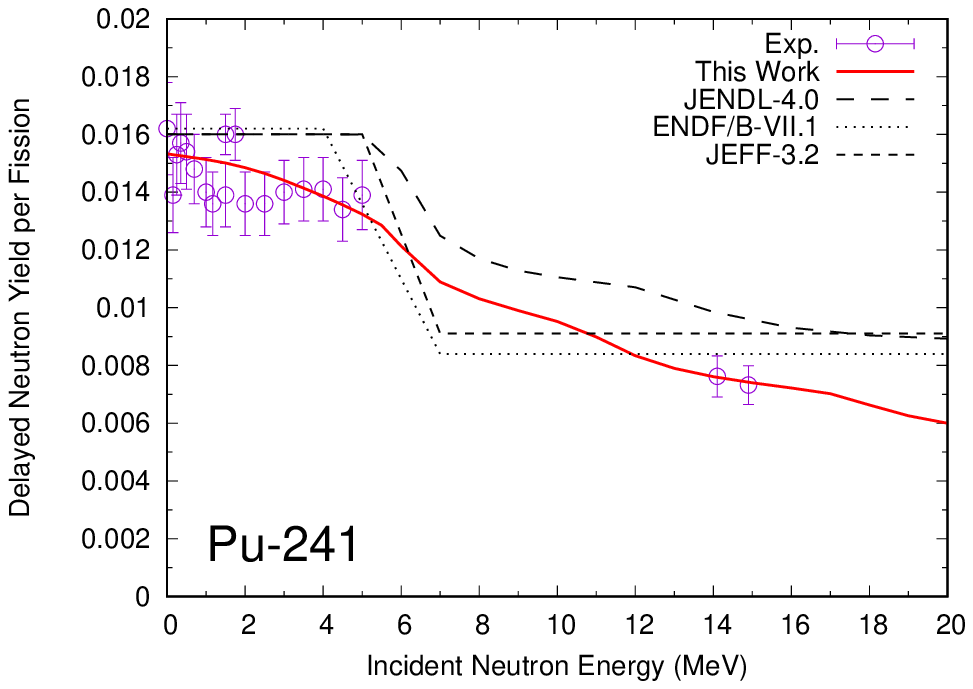}
\includegraphics[width=0.40\linewidth]{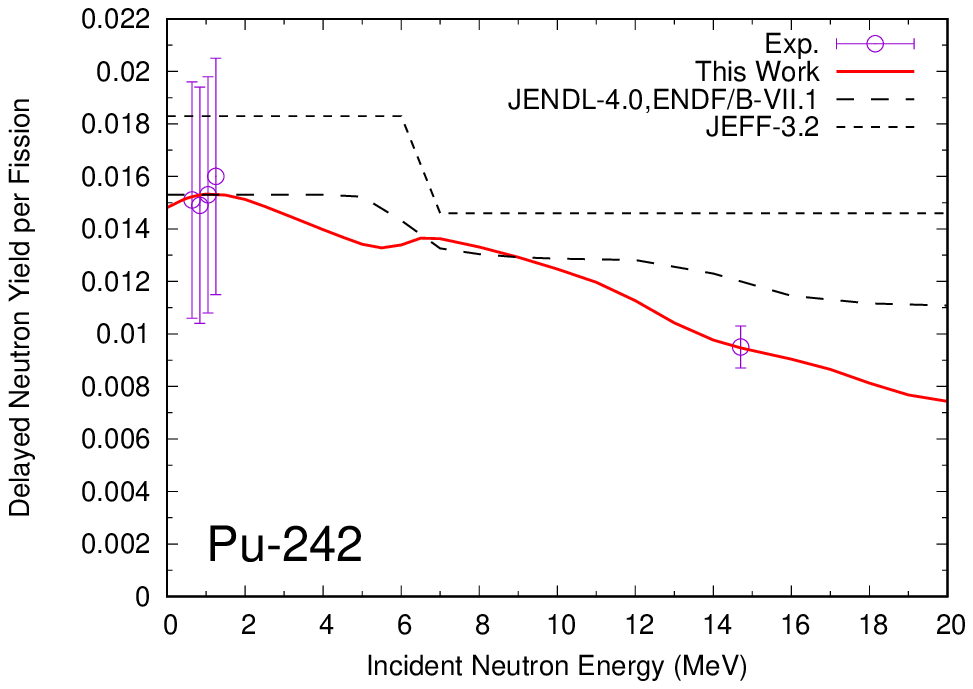}
\caption{Delayed neutron yield of plutonium isotopes from $A=239$ to $242$.}
\label{dnyplu}
\end{figure}

Figure~\ref{dnyplu} shows the result of plutonium isotopes. The experimental data of $^{239-242}$Pu are again reproduced reasonably well with our parameters.  Some lines of the evaluated data do not always pass through experimental data. This is partly attributed to the fact that the update has not been carried out for years. Like uranium isotopes, this work shows a monotonic decrease above about $4$ MeV. In the case of $^{239}$Pu, the evaluated data of JENDL-4.0 and JEFF-3.2 are approximately in a good agreement with this work up to $2$ MeV. However, deviations can be seen from $E_n>2$ MeV. This is because this work considers the experimental data of Piksaikin et al.~\cite{Piksaikin2006} (from 0.34 to 4.90 MeV) but the evaluated data do not. It should be mentioned that JEFF-3.2 largely underestimates the delayed neutron yields. For $^{240}$Pu, JENDL-4.0 looks similar to this work up to $12$ MeV, but starts to deviate around where the third chance fission occurs, because JENDL-4.0 does not consider the experimental data around $14$ MeV measured by Keepin et al.~\cite{Keepin1969,Tuttle1979}. The same occurs for delayed neutron yields at $14$ MeV in the case of $^{241}$Pu. In addition, this work considers the experimental data of Benedetti et al.~\cite{Benedetti1982}, Meadows~\cite{Meadows1976} and Cox~\cite{Cox1961,Tuttle1979}, while the evaluated data of JENDL-4.0 (and presumably JEFF-3.2 and ENDF/B-VII.1) adopts the experimental data only of Benedetti et al. Therefore, the delayed neutron yields start differently from $E_n=0$ MeV.  For $^{242}$Pu, JENDL-4.0 (ENDF/B-VII.1 adopts JENDL-4.0 evaluation) is close to this work up to $10$ MeV. It deviates after $10$ MeV because this work considers the experimental data at $14.7$ MeV of Bobkov et al.~\cite{Bobkov1989}, but JENDL-4.0 does not. JEFF-3.2 predicts delayed neutron yields larger than the others for $^{242}$Pu.

\subsection{Energy removal by pre-fission neutron}
\label{dny2}

\begin{figure}
\includegraphics[width=0.40\linewidth]{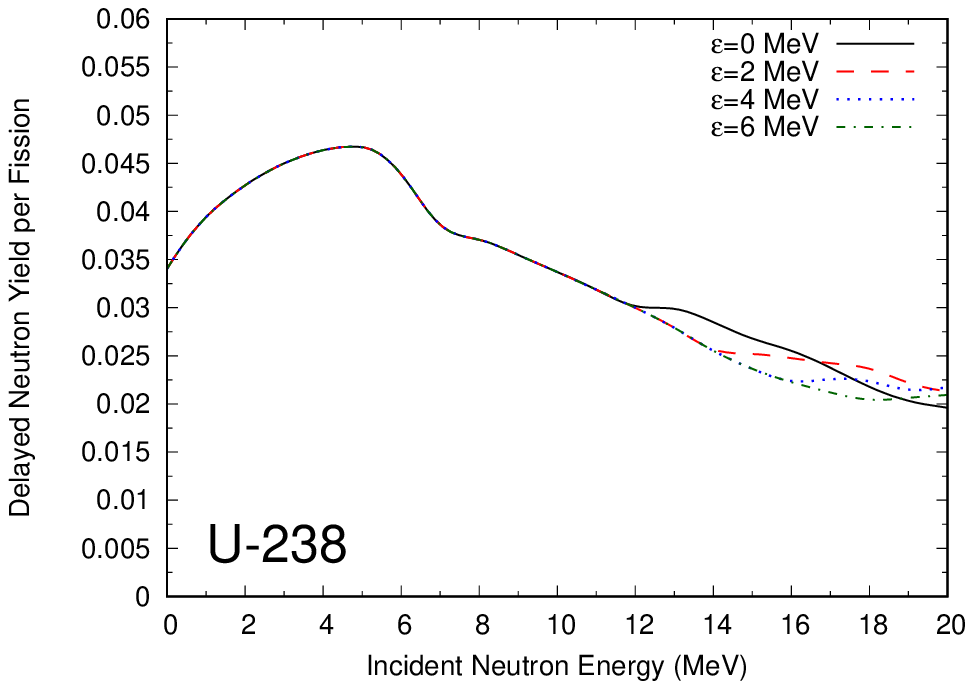}
\includegraphics[width=0.40\linewidth]{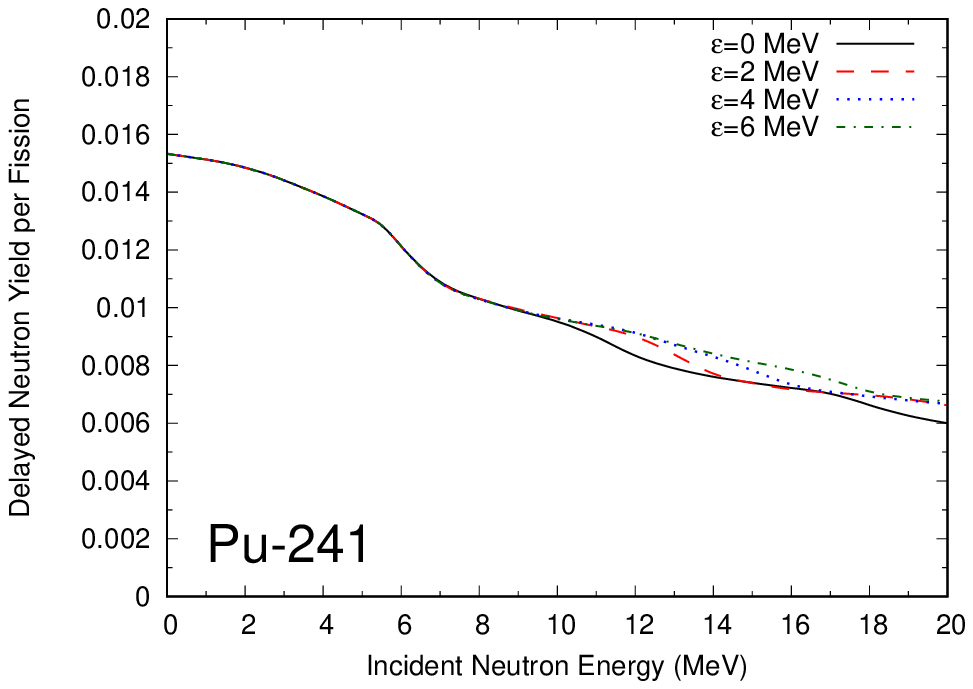}
\caption{Delayed neutron yields of $^{238}$U and $^{241}$Pu as a function of incident neutron energy with different pre-fission neutron energies. We have assumed $\epsilon_n\equiv\epsilon_n'=\epsilon_n''$. The lines in the figure are the spline curves.}
\label{epsilon}
\end{figure}

In the previous section, the contribution of the second and third chance fissions were taken into account by Eq. ~\eqref{mcfission}. However, we neglected the effect of the pre-fission neutron energy by approximating $\epsilon_n'=\epsilon_n''=0$ MeV. In order to estimate the effect of finite pre-fission neutron energy, we calculate the delayed neutron yields of $^{238}$U and $^{241}$Pu varying $\epsilon_n'$ and $\epsilon_n''$ in Eq.~\eqref{mcfission}. We simply approximate $\epsilon_n\equiv\epsilon_n'=\epsilon_n''$. If $E_x'<\epsilon_n'(E_x''<\epsilon_n'')$, we set $E_x'=\epsilon_n'(E_x''=\epsilon_n'')$. The result is shown in Fig. \ref{epsilon} changing $\epsilon_n=0$ to $6$ MeV. As the pre-fission neutron energy increases, the delayed neutron yields show different curves for both nuclides above $E_n=10$ MeV. The effect of the finite pre-fission neutron energy is not unique for $^{238}$U and $^{241}$Pu because many factors such as fission cross sections, delayed neutron yields by multi-chance fission, etc. are involved. However, we see about 10\% variations of the delayed neutron yields at most for both nuclides. The same result is obtained in other nuclides.

The effect of energy removals by pre-fission neutrons may change the parameters listed in Table~\ref{parameters}. However, we expect the delayed neutron yields obtained by the least square fitting in the previous section would not change significantly because of rearrangement of the parameters.

\section{Assessment of fission yields at thermal and fast neutron fissions}
\subsection{Comparison of Fission Product Yields}
\label{comparison}

We reproduced the delayed neutron yields reasonably well by the phenomenological parameters introduced in the previous section. The fission yields of nuclides calculated in this way, especially of delayed neutron precursors, are thus optimized to the delayed neutron yield. Therefore, we should check the validity of fission yields obtained in the previous section by comparing with experimental data.

Because most of delayed neutron precursors are not able to be measured directly from experiment, it is practical to compare our result with the evaluated data. In this work, we adopt JENDL/FPY-2011 for comparison. Since it is quite difficult to systematically compare all of the compiled fission fragments in JENDL/FPY-2011, we chose the ten most important precursors contributing to the delayed neutron yields for discussion. The ten precursors are determined by the sensitivity analysis performed in Ref.~\cite{Minato2015}. We investigated the thermal neutron fissions of $^{235}$U and $^{239}$Pu, and the fast neutron fissions of $^{238}$U and $^{239}$Pu.

\begin{figure}
\includegraphics[width=0.40\linewidth]{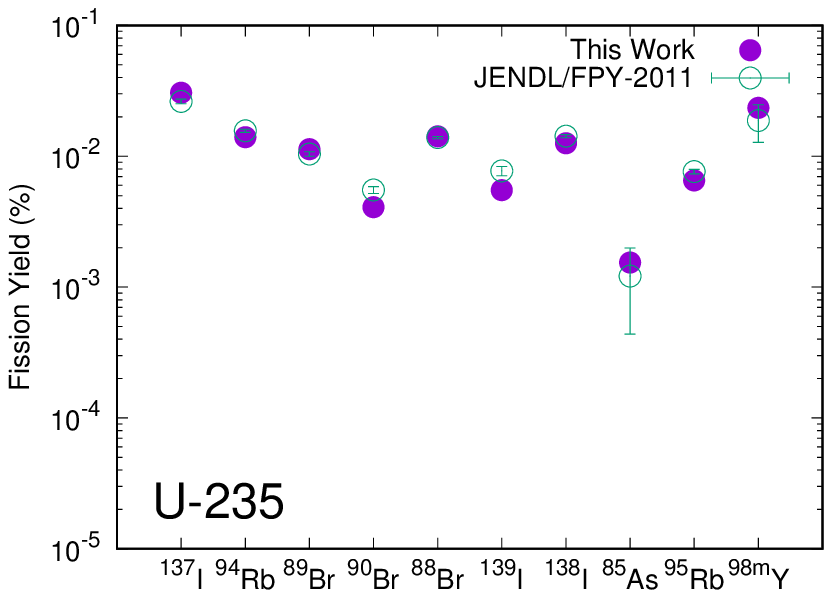}
\includegraphics[width=0.40\linewidth]{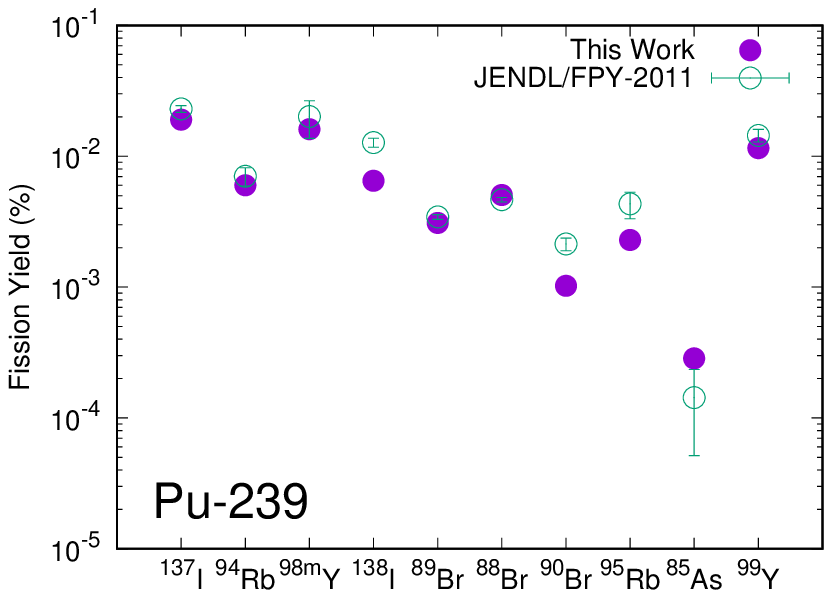}
\caption{Fission yields of ten most important precursors contributing the delayed neutron yields for $^{235}$U and $^{239}$Pu at thermal fission. The precursors are arranged in order of importance from left to right.}
\label{yieldcompthermal}
\end{figure}

\begin{figure}
\includegraphics[width=0.40\linewidth]{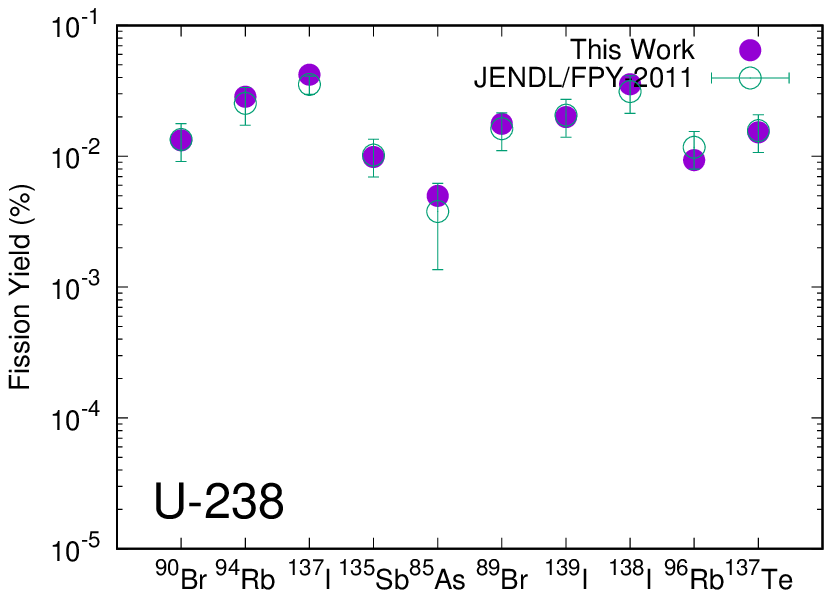}
\includegraphics[width=0.40\linewidth]{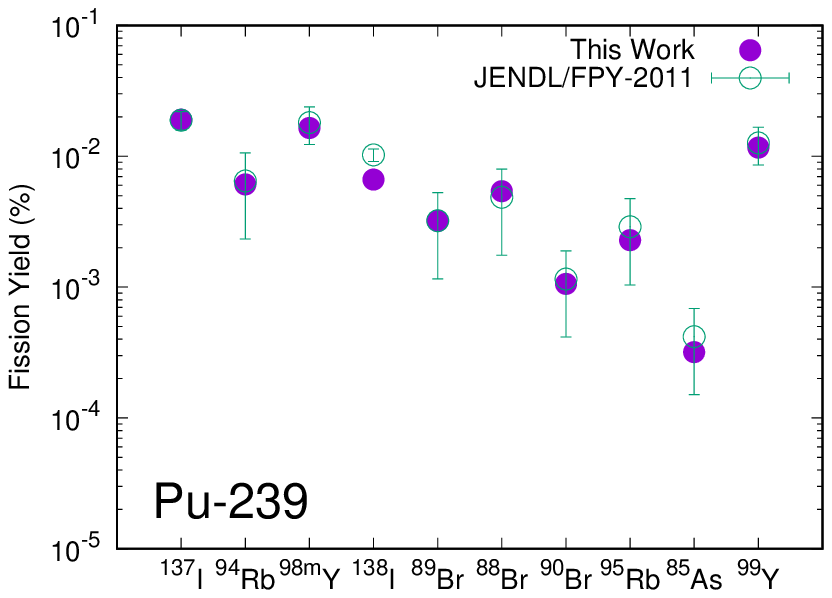}
\caption{Fission yields of ten most important precursors contributing the delayed neutron yields for $^{238}$U and $^{239}$Pu at fast fission. The precursors are arranged in order of importance from left to right.}
\label{yieldcompfast}
\end{figure}

The results for the thermal and fast neutron fissions are shown in Figs.~\ref{yieldcompthermal} and~\ref{yieldcompfast}, where the precursors are listed in order of importance from left to right. The fission yields of this work agree with JENDL/FPY-2011 within a factor of about $2$ for the ten precursors. This suggests that we do not use unnatural fission product yields that could generate inconsistencies with experimentally measured fission yields.

In order to see a difference between this work and JENDL/FPY-2011 for wide range of nuclides, we also calculated the root mean square deviations of the fission yields defined by
\begin{equation}
\sigma_{rms}=\exp\left(\sqrt{\frac{1}{N}\sum_{i}^N\left[\ln\frac{y_i}{y_i[\rm{FPY}]}\right]^2}\right),
\end{equation}
where $y_i[\rm{FPY}]$ are the fission yields of JENDL/FPY-2011 and $N$ the number of nuclides. For thermal neutron fission yields of the ten precursors, we obtain $\sigma_{rms}=1.22$ and $1.57$ for $^{235}$U and $^{239}$Pu, respectively. If we include all the nuclides, we obtain $\sigma_{rms}=3.53$ and $1.72$ for $^{235}$U and $^{239}$Pu, respectively. We thus need to keep in mind that the difference from JENDL/FPY-2011 for the other nuclides is larger than that of the ten precursors. The same is true for fast neutron fission yield. In case of the ten precursors, we obtain $\sigma_{rms}=1.16$ and $1.18$ for $^{238}$U and $^{239}$Pu, respectively. If we include all the nuclide, we obtain $\sigma_{rms}=6.69$ and $1.75$ for $^{238}$U and $^{239}$Pu, respectively. 

We list in~Table \ref{dnyield} the delayed neutron yields calculated with the thermal and fast neutron fission yield data of this work and JENDL/FPY-2011 together with the experimental data. This work reproduces the experimental data well since the parameters are adjusted to them. JENDL/FPY-2011 systematically overestimates the experimental values for thermal neutron fission, while it is close to the experimental data, as this work, for fast neutron fission.
\begin{table}
\caption{Delayed neutron yields of the thermal and fast neutron fissions calculated with JENDL/FPY-2011 \cite{Katakura2011}, this work and the experimental data are listed.}
\begin{tabular}{|c|c|c|c|}
\hline
           & JENDL/ &                 &  \\
Nuclide& FPY-2011&This Work  &Experiment\\
\hline
$^{235}$U (thermal) &  $0.01683$  & $0.01603$  & $0.0158\pm0.0005$ \cite{Keepin1957} \\
                              &                   &                  & $0.0156\pm0.0010$ \cite{Synetos1979}\\
                              &                   &                  & $0.0163$ \cite{Reeder1983}\\
                              &                   &                  & $0.0159\pm0.0004$ \cite{Saleh1997}\\
$^{239}$Pu (thermal) & $0.00781$ & $0.00625$ & $0.0061\pm0.0003$ \cite{Keepin1957}\\
$^{238}$U (fast)        & $0.04263$ & $0.04402$ & $0.0448\pm0.0042$ \cite{Keepin1957} \\
$^{239}$Pu (fast)      & $0.00691$ & $0.00658$ & $0.0063\pm0.0003$ \cite{Keepin1957} \\
\hline
\end{tabular}
\label{dnyield}
\end{table}

\subsection{Delayed Neutron Activity and Decay Heat}
\label{ieyd}

\begin{figure}
\includegraphics[width=0.40\linewidth]{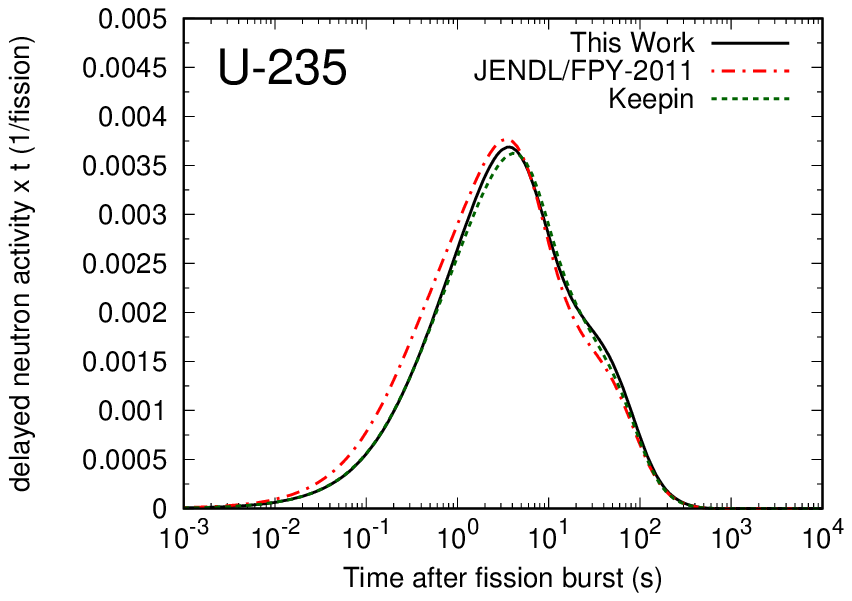}
\includegraphics[width=0.40\linewidth]{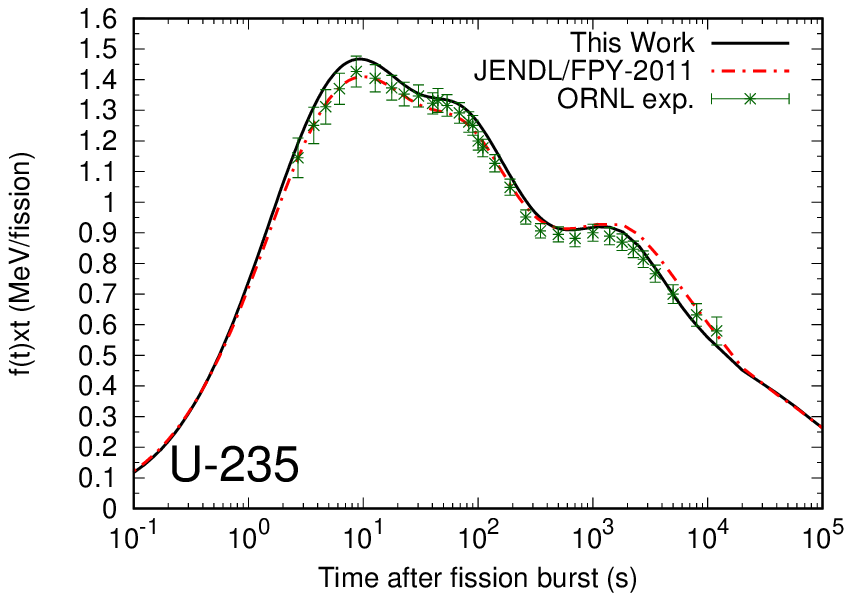}
\includegraphics[width=0.40\linewidth]{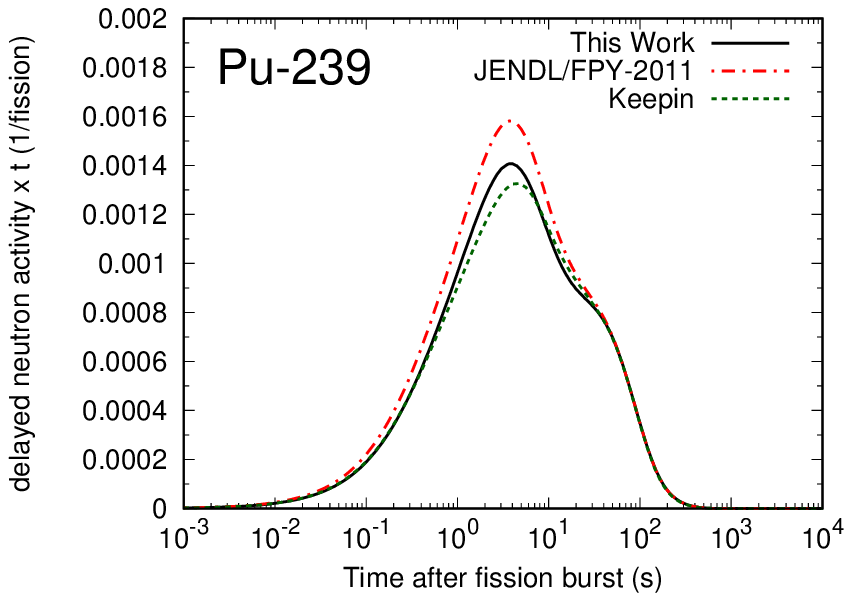}
\includegraphics[width=0.40\linewidth]{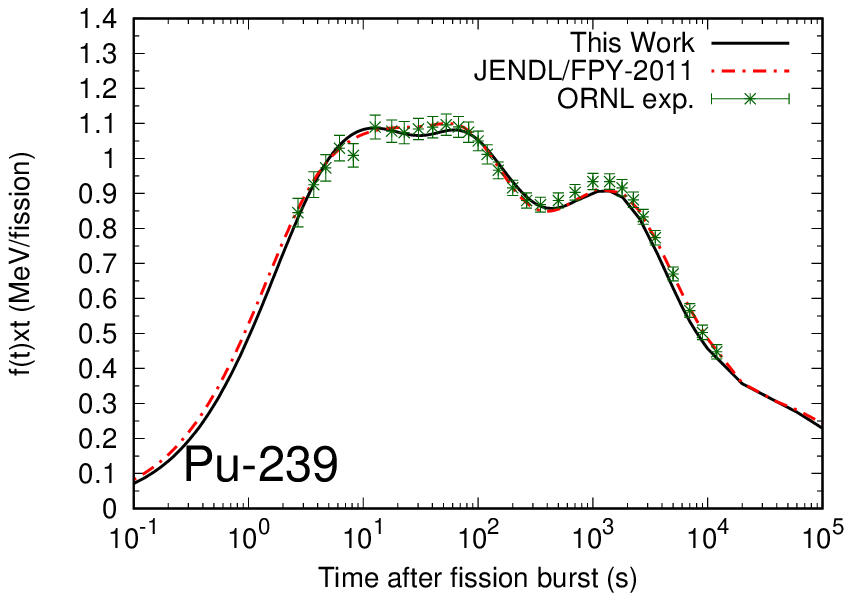}
\caption{Decay heats (the right panels) and delayed neutron activities (the left panels) multiplied by time $t$ for thermal neutron fission of $^{235}$U and $^{239}$Pu (bottom). We assume instant neutron radiation. The result of Keepin's six group \cite{Keepin1957} are also shown with the delayed neutron activity. The experimental data for the decay heat are taken from Refs.~\cite{ORNL1,ORNL2}.}
\label{final1}
\end{figure}

The fission yields affect not only delayed neutron yield but also decay heat~\cite{Katakura2011}. It is thus important to check them simultaneously to guarantee the accuracy of the yield data. Figure~\ref{final1} shows the delayed neutron activities (the left panels) and the decay heat (the right panels) for thermal neutron fissions of $^{235}$U and $^{239}$Pu. We can see that the delayed neutron activities of JENDL/FPY-2011 overestimate those of Keepin's six group evaluation ~\cite{Keepin1957}, but this work is rather close to it. Interestingly, the decay heats of this work reproduce the experimental data well, despite not being adjusted to do that. We checked, by replacing fission yields of the ten most important precursors of this work with JENDL/FPY-2011, that better agreements with the Keepin's data in this work are mainly attributed to the reduction of fission yields of $^{90}$Br ($T_{1/2}=$1.92 s), $^{94}$Rb (2.7 s), and $^{139}$I (2.280 s), which are listed in Fig.~\ref{yieldcompthermal}. Similarly, in the case of $^{239}$Pu, it is attributed to $^{94}$Rb ($T_{1/2}=2.7$ s), $^{98m}$Y (2.0 s), $^{89}$Br (4.357 s), and $^{90}$Br (1.92 s).

\begin{figure}
\includegraphics[width=0.40\linewidth]{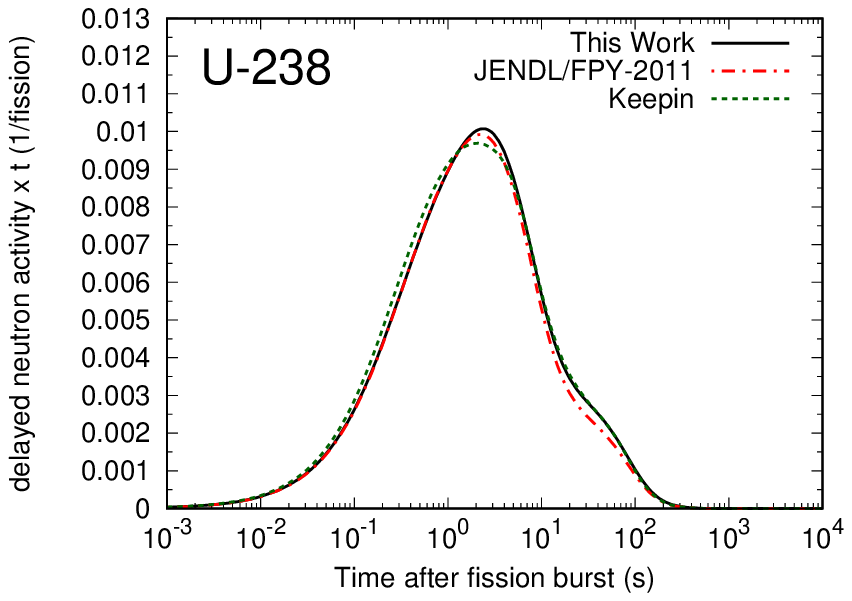}
\includegraphics[width=0.40\linewidth]{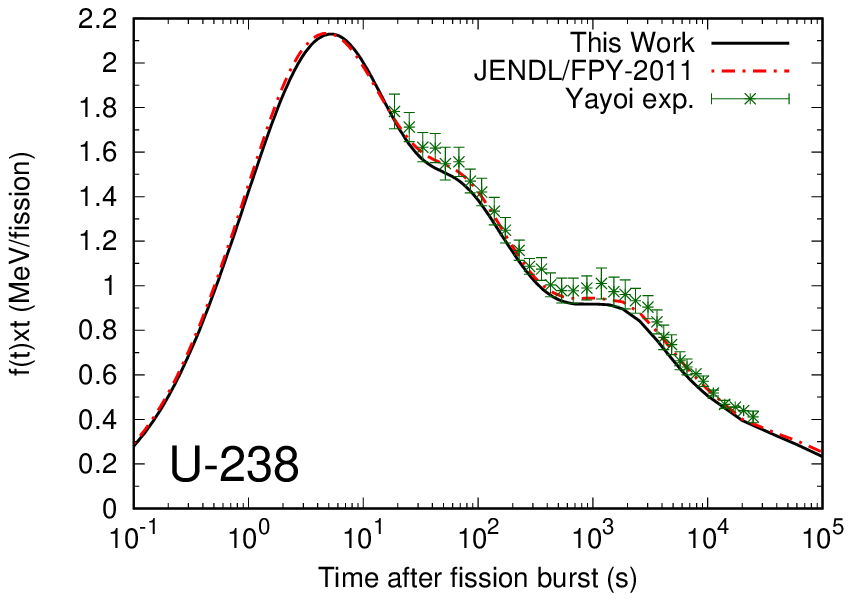}
\includegraphics[width=0.40\linewidth]{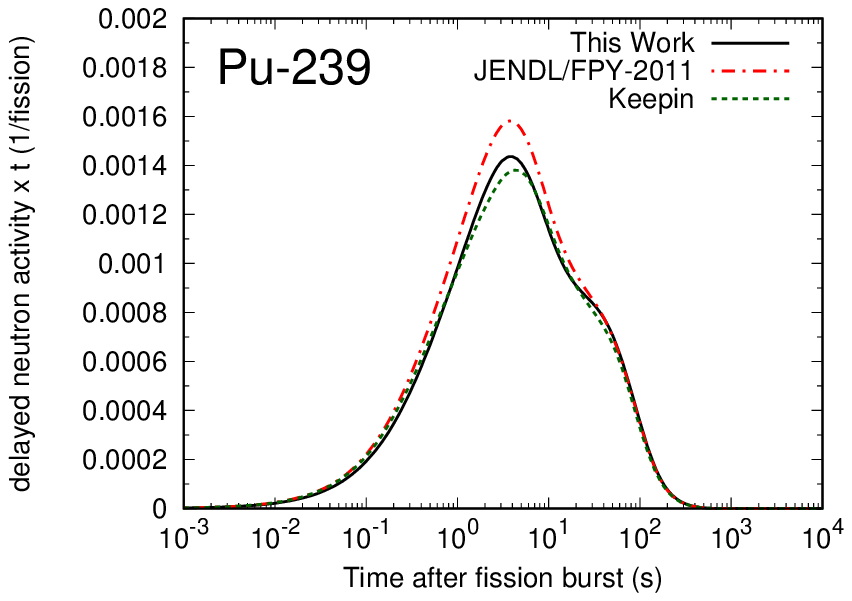}
\includegraphics[width=0.40\linewidth]{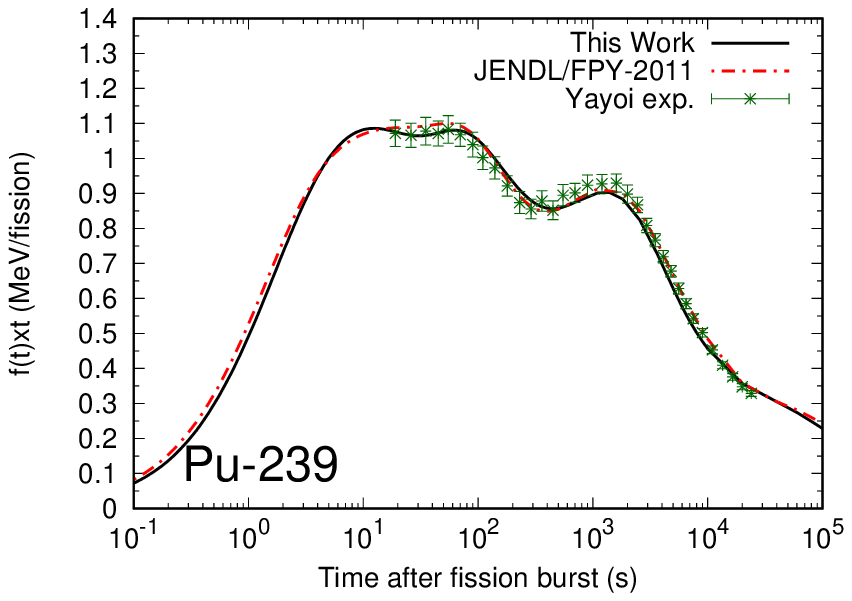}
\caption{Same as Fig. \ref{final1}, but for fast neutron fission of $^{238}$U and $^{239}$Pu. The experimental data for the decay heat are taken from Refs.~\cite{Yayoi1,Yayoi2,Yayoi3}.}
\label{final2}
\end{figure}

Figure~\ref{final2} shows the results for the fast neutron fission of $^{238}$U and $^{239}$Pu. The delayed neutron activity of this work is close to Keepin's data. JENDL/FPY-2011 also reasonably reproduces results for $^{238}$U but overestimates Keepin's data for $^{239}$Pu. Similar to the results in the thermal neutron fission case, the decay heats of this work reproduce the experimental data well although we did not consider them when fitting. 
We checked, in the same way as thermal neutron fission, that $^{138}$I ($T_{1/2}=6.23$ s) and $^{85}$As($2.02$ s) are mainly responsible for improving the delayed neutron activity of this work, compared to JENDL/FPY-2011 in the case of $^{239}$Pu.

\section{Summary}
\label{summary}
The incident neutron energy dependence of the delayed neutron yields of uranium and plutonium isotopes are calculated by the summation calculation. We introduced phenomenological parameters in the most probable charge and the odd-even effect, which were determined by least square fitting with available experimental data.

The incident neutron energy dependence of the delayed neutron yields was reproduced reasonably well with the parameters obtained by the least square fitting. However, deviations from experimental data were found for $^{233,235}$U around $E_n=4$ MeV. We discussed that the measurement of delayed neutron yield for $^{232,234}$U as well as the multi-fission cross sections of $^{233,235}$U would improve the present result further. We also compared the obtained fission yields with the evaluated fission yield data of JENDL/FPY-2011, and confirmed that the present result did not show significant deviations from the JENDL/FPY-2011, at least for ten delayed neutron precursors.

The obtained fission yields were also assessed by comparing with delayed neutron activities and decay heats. Although the fission yields obtained in this work are optimized to the delayed neutron yields, the calculated decay heats show good agreement with the experimental data. 
We tried to specify major nuclides responsible for good agreements with the experimental delayed neutron data in this work compared to JENDL/FPY-2011, by replacing the ten most important precursors in this work with JENDL/FPY-2011, and found that only a few precursors affect improvements of the delayed neutron yields. This finding will be useful for the evaluation of the next JENDL nuclear data.

Chiba recently pointed out~\cite{Chiba2016} that the correction of delayed neutron precursors does not significantly affect the decay heat by using a sensitivity calculation. It is then expected that better results on delayed neutron yields and activities can be obtained just by modifying fission yields of some important precursors of JENDL/FPY-2011. This work was able to specify important precursors for delayed neutron activities. As a next step, we plan to generate fission yield data which are able to reproduce decay heats and delayed neutron simultaneously by using the present approach and fission yield evaluations.


\subsection*{Acknowledgment}
The author thanks a great support from IAEA CRP on a Reference Database for Beta-Delayed Neutron Emission and its members. He also thanks NSCL/FRIB computing resources. This work was supported by JSPS KAKENHI Grant No. 25400303.

\end{document}